\documentclass[11pt]{article}

\usepackage[final]{acl}
\usepackage{booktabs}
\usepackage{amssymb}
\usepackage{times}
\usepackage{latexsym}
\usepackage{amsmath}
\usepackage{autobreak}
\usepackage{float} 
\usepackage{algorithm}
\usepackage{algorithmic}
\usepackage{amsmath}
\usepackage{bm}
\usepackage{arydshln}
\usepackage[T1]{fontenc}
\usepackage{multirow}
\usepackage[utf8]{inputenc}

\usepackage{microtype}

\usepackage{inconsolata}

\usepackage{graphicx}
\usepackage{booktabs}

\makeatletter
\newcommand{\rmnum}[1]{\romannumeral #1}
\newcommand{\Rmnum}[1]{\expandafter\@slowromancap\romannumeral #1@}
\newcommand{\authornote}[1]{\thanks{#1}}
\newcommand{\email}[1]{{\small\ttfamily #1}}
\makeatother

\title{Checking Fact with Better Retrieval: Dynamic Contrastive Learning for Evidence Retrieval}

\author{Zhongtian Hua\textsuperscript{1}, Yi Luo\textsuperscript{1},Meijia Yu\textsuperscript{2}, Yingjie Han\textsuperscript{1}\authornote{*}\\
       \textsuperscript{1}Zhengzhou University, Zhengzhou, Henan, China\\
       \textsuperscript{2}Henan University of Science and Technology, Luoyang, Henan, China\\ 
\email{hzt1113@gs.zzu.edu.cn}\\
\email{nancetide@stu.zzu.edu.cn}\\
\email{ieyjhan@zzu.edu.cn}\\
\email{240320261531@stu.haust.edu.cn}\\}

\begin{document}
\maketitle
\begin{abstract}
In the field of multimodal fact checking, the accuracy of retrieving evidence from different modalities has a significant impact on the downstream claim verification process. Existing general multimodal retrieval methods are often constructed based on semantics, resulting in the retrieved evidence being similar but not relevant to the claim. This paper proposes a  \textbf{D}ynamic \textbf{A}daptive \textbf{C}ontrastive \textbf{L}earning method for evidence \textbf{R}etrieval called DACLR  to address these issues. DACLR first uses a Multimodal Large Language  Model (MLLM) to uniformly convert multimodal evidence and claims into text modalities, and extracts the features of these information at event level. Then, it conducts evidence retrieval through a two-stage retrieval method of recall-rerank. DACLR enhances the  model's event perception ability of the retrieval stage by optimizing the contrastive loss and mining hard negative samples. Specifically, DACLR designs three loss functions at two levels (semantic and event) based on the InfoNCE loss.Corresponding to these, three sets of hard negative sample candidates are set up. The model dynamically adjusts the ratio based on the accuracy supervision signal of intra-batch samples, allowing the model to learn the correlation between claims and positive samples at the event level without forgetting the semantic retrieval ability. Extensive comparison and ablation experiments demonstrates the effectiveness of DACLR and its internal optimization methods. Further research also prove the advantages of DACLR in the field of multimodal evidence retrieval.
\end{abstract}

\section{Introduction}
\begin{figure}[t]
  \includegraphics[width=\columnwidth]{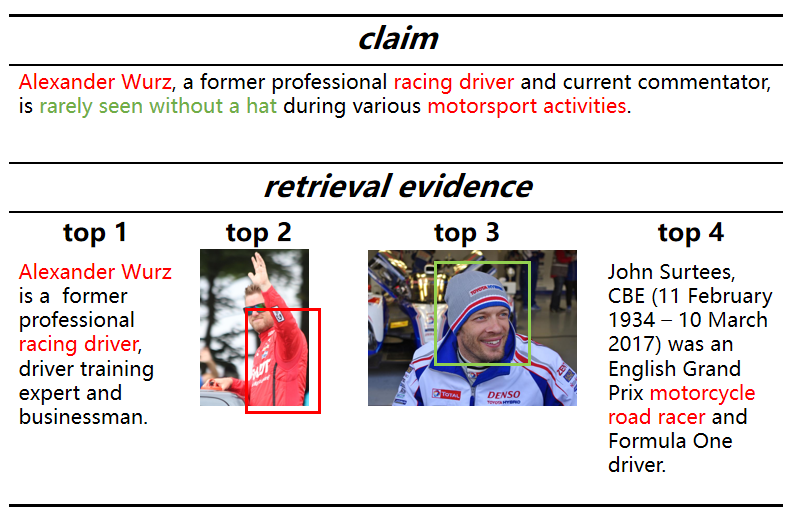}
  \caption{An example of semantic-based multimodal evidence retrieval: It presents claim and the top 4 evidence retrieved by semantic model. Although the first two pieces of evidence are highly relevant to the claim at the semantic level, the evidence that provides the most relevant information is placed at the lower position.}
  \label{fig:intro}
\end{figure}

Multimodal fact-checking aims to search for and integrate multimodal evidence (e.g., text, tables, images) from the specified multimodal evidence database to verify the veracity of the given text\cite{yang-etal-2018-hotpotqa}. Current research\cite{zhou-etal-2019-gear} typically follows a two-stage paradigm of evidence retrieval and claim verification. In the evidence retrieval stage, fine-grained evidence related to the claim is retrieved from a large-scale evidence database\cite{10.1162/tacl_a_00454}. In the claim verification stage, the authenticity of the claim is determined based on the retrieved evidence\cite{Wu_Rao_Sun_He_2021}. Existing studies\cite{zheng-etal-2024-evidence} have demonstrated the significance of the evidence selected in the retrieval stage for the entire process. The more accurate the information related to the claim in the retrieved evidence is, the more accurate the judgment made by the verification system will be.

Although the current retrieval solutions for the field of fact checking have made significant progress\cite{zhou-etal-2019-gear,zheng-etal-2024-evidence,Wu_Rao_Sun_He_2021}, most of the methods focus on text modal retrieval. On the other hand, multimodal retrieval solutions\cite{zeng-etal-2025-enhancing,zhou-etal-2024-marvel} for general domains usually construct the retrieval scheme based on semantic similarity. In this scheme, all candidate multimodal evidence and claims are mapped to the same latent semantic space by an embedding model, and then the top k candidate evidence is selected based on the similarity between the features. One common issue that arises in the practical application of these methods is that the evidence providing the most relevant information for claim verification does not rank high enough, resulting in insufficient information in the subsequent verification stage. Figure 1 shows the 4 results returned by the semantic retrieval model for the statement to be verified. It can be seen that the first two pieces of evidence are highly similar in semantics to the statement, but both lack the key fact of "wearing a hat" at the event level, while the evidence truly containing this event ranks much lower. Thus, although these models may perform well in general multimodal retrieval, they are difficult to directly handle the task of fact checking.The fundamental cause of the aforementioned problem lies in the fact that the model performs feature extraction of multimodal information during training and conducts semantic learning based on this. Semantic feature extraction often relies on semantic feature information such as participants and attributes, which tends to overlook the feature information at the event skeleton level, thereby resulting in retrieval results that are similar but not relevant\cite{10.1145/3539618.3591642}.

To address the aforementioned challenges, this paper proposes a multimodal false information retrieval method based on dynamic adaptive contrastive learning, named DACLR (\textbf{D}ynamic \textbf{A}daptive \textbf{C}ontrastive \textbf{L}earning for \textbf{R}etrieval). DACLR is divided into three main parts: event summary generation, two-stage evidence retrieval, and model training optimization. In the event summary generation part, to map claims and evidence to a unified event level, DACLR uses a predefined prompt to guide the Multimodal Large Language Model(MLLM) to generate event summaries containing event skeletons, participants, and attributes for claims and evidence. In the two-stage evidence retrieval, the process is divided into two parts: recall and re-ranking. The bi-encoder\cite{10.1145/3486250} is used in the retrieval process to improve the computational efficiency, while the cross-encoder\cite{zhang2022adversarialretrieverrankerdensetext} is used in the re-ranking stage to obtain better results. Finally, to enable the encoder to better capture the connection between claims and positive evidence at the event level, DACLR designs a dynamic adaptive contrastive learning method. DACLR uses the in-batch accuracy of the model during training as the supervision signal, thereby adjusting the loss functions and hard negative samples at the event and semantic levels . By setting the contrastive loss and negative samples for the current stage based on the model's own training situation, the model can learn the input features at the event level while not forgetting the learning content at the semantic level. In summary, the contributions of this paper can be summarized as:
\begin{itemize}
\item[$\bullet$] A multimodal evidence retrieval method named DACLR based on dynamic adaptive contrastive learning is proposed to achieve more accurate multimodal retrieval.
\item[$\bullet$]To enable the retrieval model to learn the intrinsic relationship between claims and evidence at the event level, a dynamic adaptive contrastive learning method is designed.
\item[$\bullet$]Experiments are conducted to prove the effectiveness of DACLR. Detailed experimental analysis shows that DACLR not only performs better than existing multimodal retrieval methods in the multimodal datasets, but also outperforms existing fact-checking retrieval methods in the single-modal datasets.
\end{itemize}
\section{Related Work}
\subsection{Retrieval for fact checking}

The mainstream paradigm of fact checking can be divided into two parts: retrieval and verification. The evidence retrieval results obtained in the retrieval stage have a significant impact on the subsequent verification. The early retrieval relied on TFIDF and BM25 \cite{10.1561/1500000019}, which encode claims and evidence into sparse bag-of-words vectors to achieve efficient word matching. However, these methods are limited by the vocabulary gap problem and are difficult to capture the semantic relationship between claims and evidence. After the emergence of transformer\cite{10.5555/3295222.3295349} and its variants, retrieval systems based on pre-trained models have become the mainstream paradigm\cite{humeau2020polyencoderstransformerarchitecturespretraining}. This method encodes queries and documents into dense semantic vectors \cite{ren-etal-2021-pair,zhang2022hlatrenhancemultistagetext}, and uses an optimized similarity search algorithm\cite{8733051} to perform retrieval. These methods require training a retrieval model using a contrastive learning paradigm or a supervised fine-tuning paradigm, thereby obtaining the evidence with the highest similarity to the claim.

However, in the field of checking, the evidence that is most similar to a claim might not be the most relevant one, and the truly relevant evidence might be ranked lower by the retrieval model due to insufficient semantic similarity. To address this issue, some studies\cite{zheng-etal-2024-evidence} have attempted to achieve joint optimization between the retriever and the verification model by introducing KL divergence as a supervisory signal. Other studies\cite{malviya-katsigiannis-2024-evidence} have utilized the interaction relationships between evidences through multi-stage reordering. However, these methods either increase the system complexity or still focus on semantic similarity in the retrieval stage, failing to fundamentally solve the retrieval bias problems of the model at the semantic and event levels. Moreover, most of the existing research is focused on pure text modal fact checking, and there is an urgent need for a unified retrieval framework for multimodal fact checking.
\begin{figure*}[]
  \includegraphics[width=\textwidth]{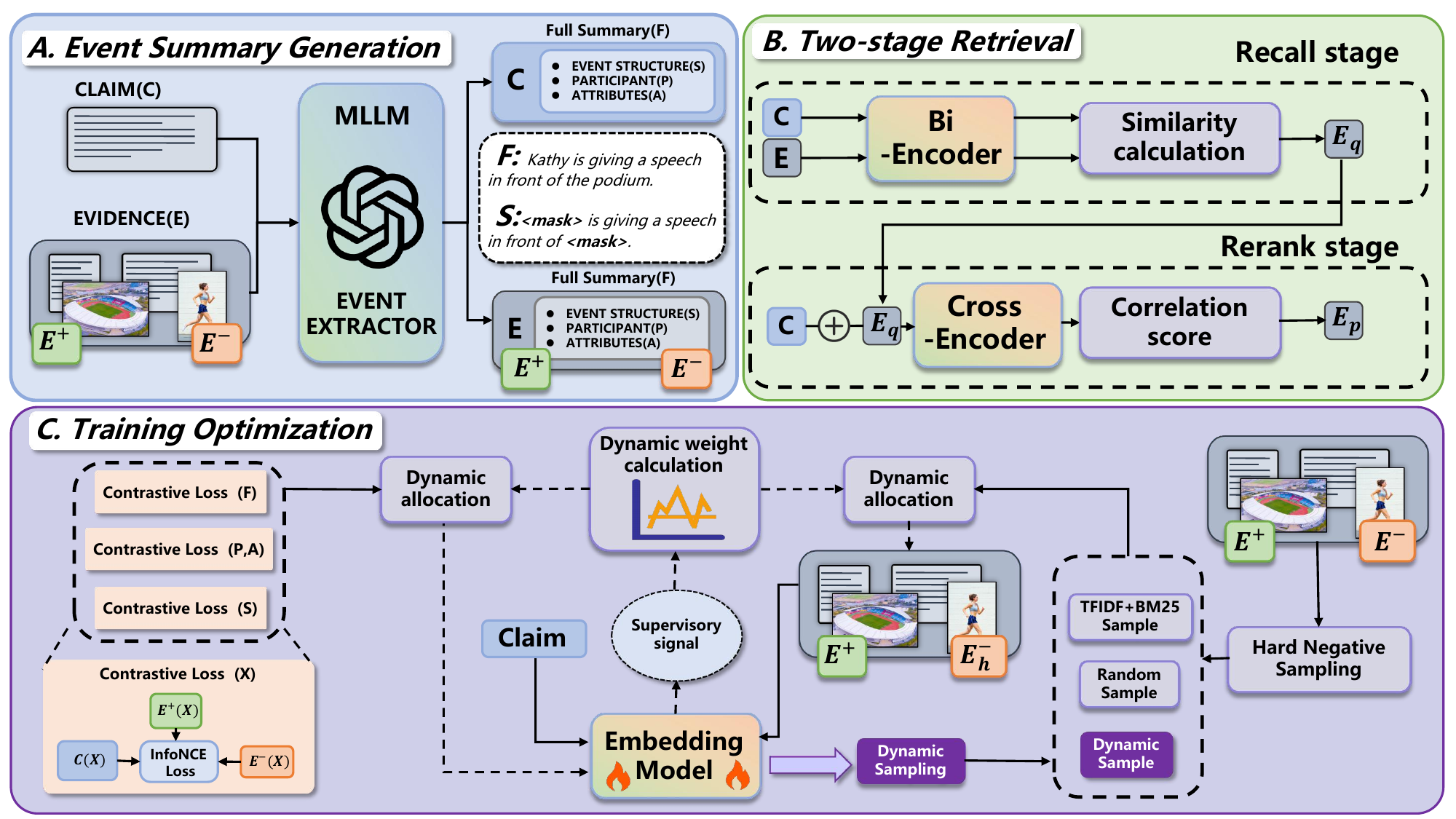}
  \caption{An overview of DACLR.}
  \label{fig:overview}
\end{figure*}
\subsection{Multimodal retrieval}
Compared with text-modal retrieval, multimodal retrieval requires the system to capture the features and connections between information from different modalities \cite{yuan2021multimodalcontrastivetrainingvisual}. A mainstream approach is to use representation models. Early studies encoded the information of the visual and the text information separately and concatenated the vectors for information fusion \cite{radford2021learningtransferablevisualmodels}. Recent research \cite{li2024improving} focuses on projection-based methods. This type of method utilizes models like CLIP \cite{radford2021learningtransferablevisualmodels} to convert visual inputs into feature sequences and introduces a projection layer to align these sequences with the text embeddings. For example, MARVEL\cite{zhou-etal-2024-marvel} uses this framework to capture multimodal information in the output space of the language model; MCL \cite{li2024improving} adds a retrieval token, and CIEA\cite{zeng-etal-2025-enhancing} captures supplementary information between text and images for better multimodal retrieval. Although these representation model-based methods can achieve effective retrieval in general domains, they are difficult to be adjusted according to the characteristics of specific downstream tasks (such as fact checking).

Another approach is to uniformly convert the information of other modalities into text modality, thereby transforming the multimodal retrieval task into text retrieval task \cite{baldrati2023zeroshotcomposedimageretrieval}. Although this approach may result in information loss during the modal conversion process \cite{che2024enhancingmultimodalunderstandingclipbased,10658022}, it is relatively easy to adapt and enhance according to the characteristics of the downstream task. The DACLR follows this method route and has built an end-to-end multimodal retrieval method for fact checking.

\section{Method}

This section provides a detailed introduction to the proposed method Dynamic Adaptive Contrastive Learning for Retrieval(DACLR). The overall architecture of the method is shown in Figure 2, which consists of three main components: Event Summary Generation, Two-stage Retrieval, and Training Optimization. Given a claim to be tested and an evidence set $E=\{e_1,e_2,……,e_n\}$, the goal of DACLR is to retrieve the top $q$ most relevant evidence from the set $E$ that are related to the claim.

\subsection{Event Summary Generation}

To transform multimodal evidence into a unified representation, MLLMs are employed to extract event summaries for each claim and multimodal evidence. In order to enhance the information alignment between evidence and claims, thereby improving the modeling of factual consistency and the sorting of retrieval results. Drawing on the discourse semantic theory (Penn Discourse Treebank, PDTB)\cite{prasad-etal-2008-penn}, the claim/evidence is decomposed into structure, participants and attributes.Specifically, for the input $x$, the model outputs an event skeleton tuple:

\begin{equation}
\begin{aligned}
&S(x)=\{summary(x),participants(x),\\
&attributes(x),structure(x)\}
\end{aligned}
\end{equation}

where summary is a natural language summary of an event, including participants, attributes, and event skeleton, participants and attributes respectively list the participant entities and their attributes in the event, structure is the remaining event skeleton after masking all the participants and attributes in the event summary with masks, such as '[Mask] is involved in [Mask]'. The prompt used in this section will be presented in the appendix.

\subsection{Two-stage Retrieval}

During the retrieval stage, the classic recall-rerank framework is followed. Among them, the objective of the recall stage is to narrow down the range of n candidate evidences to $p$ (where $q < p < n$). At this stage, the claim and evidence will be independently encoded by the bidirectional encoder of DACLR, and the correlation between vectors will be calculated through cosine similarity. The top $p$ pieces of evidence with the highest scores are selected.
Given the claim and set $E$, the bidirectional Encoder will encode them to obtain their respective representations:

\begin{equation}
\begin{aligned}
f_c,f_e = Encoder(c,e)
\end{aligned}
\end{equation}

Next, for each claim, calculate the pairwise cosine similarity Sn between it and all the evidence in the evidence set:

\begin{equation}
\begin{aligned}
&S_n = \{s_1,s_2,……,s_n\} = \{cos(f_c,f_{e_1}),\\
&cos(f_c,f_{e_2}),……cos(f_c,f_{e_n})\}
\end{aligned}
\end{equation}

Sort $S_n$, and the top $p$ pieces of evidence will be input into the rerank stage as the final result $s_p$ of the recall stage.

The goal of the rerank stage is to select the most relevant $q$ pieces of evidence from the $p$ pieces of $s_p$. In this stage, DACLR refers to the implementation of the cross-encoder and concatenates the claim and evidence together for feature encoding, thereby obtaining the joint representation of the claim and the evidence. A fully connected layer is used to obtain the score for each piece of evidence in the $s_p$:

\begin{equation}
\begin{aligned}
 &\widetilde{S}_p = \{s_1,……,s_p\} =\{FC(Encoder(c \oplus e_1)),\\
 &……,FC(Encoder(c \oplus e_p))\}
\end{aligned}
\end{equation}

Sort $S_p$ and ultimately select $q$ pieces of evidence as the final results.

\subsection{Training Optimization}

In order to enable encoder model to better learn the relationship between claims and positive sample evidence at the event level, DACLR uses a dynamic adaptive contrastive learning method. This method optimizes the two major elements of contrastive learning: contrastive loss and hard negative sample mining, and uses the in-batch accuracy of the model as a self supervised signal to dynamically regulate the training process.

\subsubsection{Dynamic loss calculation}

DACLR designs two layers of contrastive loss, namely the event summary loss and the semantic-event joint loss. Among them, the event summary loss is calculated using the event summaries, which includes information from both the semantic-level and event-level. Therefore, it participates in the entire training process and holds a dominant position in the training. The semantic-event joint loss is dynamically matched with the proportion of semantic(participants and attributes) and event(skeletons) in the training based on the self supervised signal of the model.

Event Summary Loss: This loss is calculated using the event summary of claim and evidence. For each claim, calculate the comparative loss:

\begin{equation}
\begin{aligned}
&L_{full}\ =\ \\
&- \log{\frac{e^{sim(f_{c_i}^{full},f_{e^+}^{full})/\tau)}}{e^{sim(f_{c}^{full},f_{e^+}^{full})/\tau)}+\sum_{D^-}{e^{sim(f_{c}^{full},f_{e^-}^{full})/\tau)}}}}
\end{aligned}
\end{equation}

where $f^{full}$ is the representation after encoding and pooling, $\tau$ is the temperature hyperparameter, and $D-$ is the currently negative sample set.

semantic-event joint loss: Furthermore, DACLR designs two sub-losses, which are combined into semantic-event joint loss by dynamic parameter adjustment:
Semantic loss $L_{sent}$: Compare the claim's participants and attributes with the corresponding version of the evidence.
Event comparison loss $L_{struct}$: Use the structure field for matching, forcing the model to focus on the structural pattern of the event.
The final semantic-event joint loss is:

\begin{equation}
\begin{aligned}
L_{unit}=\beta \cdot L_{sent}+(1-\beta) \cdot L_{struct}
\end{aligned}
\end{equation}

where $\beta$ is a dynamic weight used to balance the contrastive loss of two layers. The calculation method of this weight is detailed in section 3.3.3.

\subsubsection{Dynamic Negative Sample Mining}

Traditional contrastive learning usually employs random negative samples or static hard negative samples, but it is difficult to train a retriever with strong discriminative ability. DACLR proposes a hybrid negative sample mining mechanism. The negative sample pool is composed of three parts: \textbf{(\rmnum{1}).Random Negative Sample $D_{rand}$}: Selected randomly from the evidence database.\textbf{(\rmnum{2}).Shallow Hard Negative Sample $D_{TB}$}: This sample is based on the integrated retrieval results of TF-IDF and BM25, and gathers evidence that is similar to the statement words but has a relatively weak semantic correlation. \textbf{(\rmnum{3}).Self-Retrieval Hard Negative Sample $D_{model}$}: Utilizing the most similar but incorrect evidence dynamically mined by the current retrieval model during training, these samples are constantly updated as the model's capability improves. DACLR refers to existing research\cite{li2024conanembeddinggeneraltextembedding} and updates based on the model's score on the negative samples. 

During the training process, for each claim, the total number of negative samples is fixed at $K$. Among them, the proportion of fixed negative samples ($D_{rand}$ and $D_{TB}$) is $1 - p_{dyn}$, and the proportion of dynamic negative samples is pdyn. The calculation method of the proportion $p_{dyn}$ is elaborated in Section 3.3.3.

\subsubsection{Adaptive Scheduling based on Discrimination}

DACLR designs an adaptive scheduling mechanism based on the current discrimination ability of the model to calculate the matching parameters $\beta$ and $p_{dyn}$. The average positive and negative sample similarity gap of each batch is defined as:

\begin{equation}
\begin{aligned}
\Delta_t = \frac{1}{B}\sum_{i\in B}\Bigl( &\mathrm{sim}(f_{c_i}^{\mathrm{full}}, f_{e_i^+}^{\mathrm{full}}) \\
&\quad - \frac{1}{D}\sum_{j\in D}\mathrm{sim}(f_{c_i}^{\mathrm{full}}, f_{e_j^-}^{\mathrm{full}})\Bigr)
\end{aligned}
\end{equation}

And use the exponential moving average method to smooth and update $\Delta_t$:

\begin{equation}
\begin{aligned}
{\bar{\mathrm{\Delta}}}_t\ =0.9{\bar{\mathrm{\Delta}}}_{t-1}+0.1\mathrm{\Delta}_t
\end{aligned}
\end{equation}

In order to smoothly map the $\Delta_t$ to a certain value within the (0, 1) interval, DACLR uses the sigmoid function for mapping, and periodically calculates the  accuracy $ACC_{val}$ on the validation set, using it as the median threshold $\Delta_{mid}$ for the sigmoid function. Finally, the dynamic negative sample ratio and the loss weight are calculated as follows:

\begin{equation}
\begin{aligned}
p_{dyn}\ =\ \sigma(\frac{{\bar{\mathrm{\Delta}}}_t\ -\ \mathrm{\Delta}_{mid}}{\tau}),\beta=1-p_{dyn}
\end{aligned}
\end{equation}

Where $\sigma$ is the sigmoid function, and $\tau$ controls the steepness of the transition. When the model's discrimination ability is weak, it mainly uses simple negative samples and focuses on semantic feature learning; when the model gradually becomes stronger, it automatically introduces more difficult negative samples and shifts to event skeleton learning. The pseudo-code of the training process is presented in appendix.

\section{Experiments}

\begin{table*}
\renewcommand{\arraystretch}{1.5}
  \centering
  \resizebox{\linewidth}{!}{
  \begin{tabular}{cccccccccccccc}
    \toprule[1.2pt]
               & \multirow{2}{*}{\textbf{Method}}  & \multicolumn{6}{c}{FEVER} & \multicolumn{6}{c}{FEVEROUS} \\
    \cmidrule(r){3-8}\cmidrule(r){9-14}
           &  &      REC@20 & REC@100 & MRR@10    &  MRR@20 & NDCG@10 & NDCG@20 &  REC@20 & REC@100 & MRR@10    &  MRR@20 & NDCG@10 & NDCG@20                    \\
    \hline

\multirow{7}{*}{\textit{Text-only}}  &   TF-IDF &   47.87     & 70.26 & 31.74 &  33.57    & 28.99 & 29.62 &  67.11 & 71.56 & 29.71  & 29.89     & 30.14  & 29.93  \\
 &   BM25   &  46.34  & 68.77 & 28.02 & 28.32   & 30.23 & 31.26 & 69.40 & 75.53 &  33.71 &  34.03  & 31.05 & 31.74 \\

& BERT &  87.47  & 90.34 & 57.23 &   58.36  & 58.12 & 59.47 &  81.54 & \textbf{89.33} & 54.17  & 55.93 & 52.49  & 53.61  \\

 & XLNet & 85.34 &  88.46 & 60.23 &   61.92 & 62.87 & 64.14 & 76.32 & 82.95 & 58.12  & 59.73  & 55.42  & 56.21 \\

 & RoBERTa  &  85.57   & 90.65  & 55.24 &  53.67  & 54.71 & 55.24 & 80.35 & 88.13 & 55.38  & 57.42  & 59.29 & 60.01  \\
 & GERE  & 77.16  & 88.75 & 59.83 &  61.22   & 61.72 & 62.33 & 84.43 & 87.62 &  66.03  &  66.84 & 67.09 & 67.93 \\
 & KGAT  & 87.47 & 89.63 & 61.73 &  63.77   & 59.88 & 60.74 & 86.04 & 87.35 &  64.79  &  66.21   & 63.05 & 64.12 \\
 & RAV  & 87.77 & 90.57 & 60.24 &  62.29   & 62.42  & 65.71 & 85.17 &88.74 & 61.37  &  63.30  & 64.15 & 64.24 \\
 & M-ReRank  & 86.13  & \textbf{90.78} & 63.27 &  63.82   & 65.34 & 67.92 & \textbf{87.02} & 88.13 & 67.21 &   67.88 & 65.19 & 66.23 \\
\cdashline{2-14}
 & DACLR  & \textbf{88.48} & 89.58 & \textbf{69.04} &  \textbf{69.32}   & \textbf{70.28} & \textbf{71.29} & 86.93 & 88.87 & \textbf{70.11} &  \textbf{70.24}  &  \textbf{68.91} & \textbf{69.03}\\

    \cline{1-14}
    
     &   & \multicolumn{6}{c}{MMCV} & \multicolumn{6}{c}{Mocheg} \\
    \cmidrule(r){3-8}\cmidrule(r){9-14}
           & &     rec@20 & rec@100 & mrr@10    &  mrr@20 & ndcg@10 & ndcg@20 &  rec@20 & rec@100 & mrr@10    &  mrr@20 & ndcg@10 & ndcg@20                    \\
    \cline{1-14}

\multirow{7}{*}{\textit{Multimodal}}     & CLIP-DPR &   55.37    &  67.36  & 40.74 &  39.50 & 35.33& 36.23 & 44.88 & 59.63 & 31.41 &  32.85   & 30.74 & 32.49  \\

 & VinVL-DPR &  50.12    & 66.78 & 33.75 & 34.74   & 34.32  & 35.66 & 49.71 & 60.11 & 33.41 &   34.71   & 31.71 & 31.84 \\
 
 & UniVL-DP & 55.24 & 68.36  & 39.63 & 41.24  & 41.25 & 40.27  & 53.74  & 60.75 & 36.21 &  36.78    & 37.84 & 39.23 \\
 
 & T5-ANCE  &   61.35 & 69.96  & 44.25 & 46.34  & 46.38  & 47.25 &  54.32 & 63.79 & 34.77 &  34.93  & 36.87 & 37.01 \\
 
 & MARVEL  & 63.46  & 79.77   & 50.76 & 50.97  & 44.35 & 48.06 & 61.74  & 65.29 & 49.23 & 50.12   & 48.13 & 48.65 \\
 
 & CIEA  & 67.70 & 80.47 & 51.75 & 52.67   & 46.73 & 45.69 & 64.23 & 70.14 & 51.12 &   51.84  & 53.73 & 54.14 \\
 
 \cdashline{2-14}
 & DACLR  & \textbf{71.16} & \textbf{80.83} & \textbf{66.26} &  \textbf{67.13}   & \textbf{62.43} & \textbf{63.91} & \textbf{67.43} & \textbf{76.87} & \textbf{60.16} & \textbf{60.67}   & \textbf{59.41} & \textbf{60.12} \\

    \bottomrule[1.2pt]
  \end{tabular}
  }
  \caption{\label{citation-guide}
    This table shows the main experimental results. 
  }
\end{table*}

\subsection{Setup}

\subsubsection{Dataset and Metrics}
To comprehensively evaluate the retrieval capabilities of DACLR in the field of fact checking, following datasets are selected:
\paragraph{Text-only dataset} FEVER\cite{Thorne18Fever}: contains 185,455 annotated claims and 5,416,537 Wikipedia documents; FEVEROUS\cite{Aly21Feverous}: contains 87,026 annotated claims from Wikipedia.
\paragraph{Multimodal dataset} MMCV\cite{wang-etal-2025-piecing}: contains 15,569 claims along with a corresponding multimodal evidence database; Mocheg\cite{yao2023endtoendmultimodalfactcheckingexplanation}:A multimodal fact-checking dataset containing 15,601 labeled statements, 33,880 pieces of textual evidence, and 12,112 images of evidence.

The detailed statistics of all datasets is provided in the appendix.

In order to compare the performance of the DACLR with that of the baseline methods, Recall@k is used as one of the metrics in the experiment. Additionally, to analyze the ranking of positive samples among all retrieved samples, MRR@k and NDCG@k is also used in the experiment. For the values of k in all the above-mentioned metrics, they are as follows: [10, 20, 100] 

\subsubsection{Baselines and Details}
Two different types of baselines are used to be compared with DACLR:  
\paragraph{Evidence retrieval for text modalities} This type of method focuses on selecting the most relevant evidence from a text database to support a claim. For instance, the \textbf{BM25/TFIDF} methods adopted by the official FEVER dataset. Recent studies usually conduct dense retrieval based on retrieval models such as \textbf{BERT}\cite{devlin-etal-2019-bert}, \textbf{XLNet}\cite{zhong-etal-2020-reasoning}, and \textbf{RoBERTa}\cite{zhong-etal-2020-reasoning}. Among them, \textbf{GERE}\cite{zhou-etal-2019-gear}, \textbf{KGAT}\cite{liu-etal-2020-fine}, \textbf{RAV}\cite{zheng-etal-2024-evidence}, and \textbf{M-ReRank}\cite{malviya-katsigiannis-2024-evidence} are used as representatives of the latest methods. 
\paragraph{Multimodal Evidence Retrieval} This type of method aims to find the evidence  from a multimodal evidence database. The latest research selected here serves as the baselines : Pre-trained multimodal model VinVL-DPR \cite{9577951} and CLIP-DPR \cite{radford2021learningtransferablevisualmodels}, Hard negative samples mining based method UniVL-DR\cite{liu2023universalvisionlanguagedenseretrieval}, Projection-based method MARVEL \cite{zhou-etal-2024-marvel} and CIEA\cite{zeng-etal-2025-enhancing}.

\paragraph{Implementation Details}The RoBERTa model is used as the base model for training.Due to page limitations, please refer to the appendix for the implementation details of all algorithms used in the experiment and the corresponding hyperparameters.

\subsection{Main Results}

Table 1 presents the comparative main results of DACLR and various baselines in the evidence retrieval task. Overall, DACLR achieves significant improvements in retrieval performance compared to existing methods. Although it does not show a clear advantage in the recall, it achieves a substantial lead in MRR and NDCG. This result is consistent with the design goal of DACLR, which is to improve the ranking quality of the most relevant evidence in the candidate set rather than simply maximizing the number of positive examples. 

In the pure \textbf{text retrieval} task, DACLR outperforms the recent representative methods RAV and M-ReRank on both the FEVER and FEVEROUS datasets. Specifically, DACLR achieves 69.04 MRR@10 and 70.28 NDCG@10 on the FEVER dataset, and 70.11 and 68.91 respectively on the FEVEROUS dataset. The above experimental results indicate that DACLR can effectively model and retrieve evidence at the event level, thereby elevating positive evidence to a more prominent position in the retrieval list.

In the \textbf{multimodal retrieval} task, DACLR also demonstrates good generalization ability. On the MMCV and Mocheg multimodal datasets, this method achieves 66.26 and 60.16 MRR@10, and 62.43 and 59.41 NDCG@10 respectively. Although the introduction of heterogeneous representation alignment difficulties for image and text in multimodal data poses challenges, DACLR can maintain relatively stable performance. The reason for the lower results in the multimodal task compared to the text task may be due to the semantic gap between the visual and text modalities, which increases the difficulty of cross-modal retrieval. Nevertheless, DACLR can effectively coordinate heterogeneous information through the unified representation of event structure, demonstrating good cross-modal retrieval potential.

\subsection{Ablation Study}

\begin{table}
\resizebox{\linewidth}{!}{
\begin{tabular}{lccc}
\toprule
Method & REC@20 & MRR@20 & NDCG@20 \\
\midrule
RoBERTa & 85.57 & 53.67 & 55.24\\
DACLR w/o Event Loss  & 87.47 & 64.12 & 65.47 \\
DACLR w/o Hard Samples  & 86.96 & 63.31 & 66.12  \\
DACLR  & \textbf{88.48} & \textbf{69.32} & \textbf{71.29}  \\
\bottomrule
\end{tabular}  
}
  \caption{\label{citation-guide}
    This table shows the ablation experimental results on FEVER Dataset. 
  }
\end{table}

The core of the DACLR design lies in the dynamic allocation of the event loss function and the mining of difficult negative samples. By separately removing the dynamic allocation loss function (i.e., only using the InfoNCE function for training) and the mining of difficult negative samples (i.e., only using random negative samples), the effects and contributions of both are verified on DACLR. The experimental results are shown in Table 2. Overall, using the DACLR method is significantly superior to the RoBERTa model trained with standard contrastive learning in all indicators. Specifically, DACLR outperforms either without the event loss or any of the hard samples, indicating the effectiveness of these two for model training optimization.

\begin{figure}[t]
  \includegraphics[width=\columnwidth]{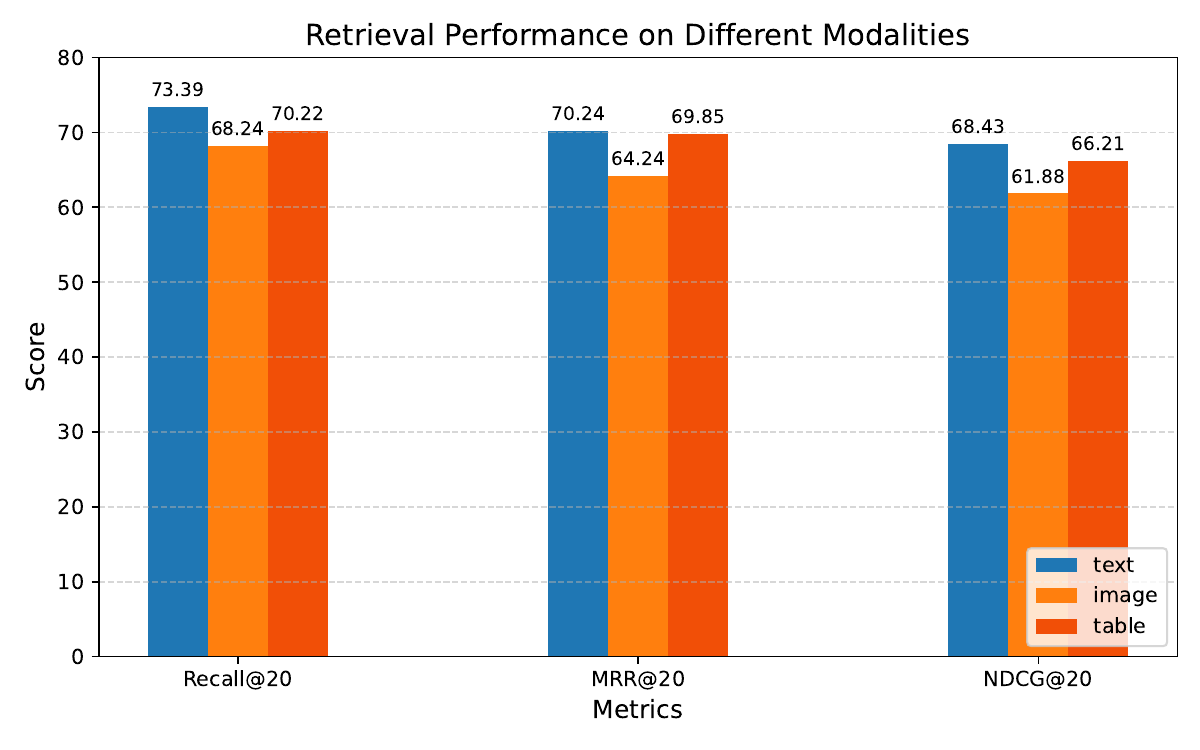}
  \caption{A figure shows the retrieval performance on different Modalities.}
  \label{fig:experiments}
\end{figure}

Furthermore, in order to comprehensively analyze the retrieval capabilities of DACLR for various modalities of evidence, experiments for each modality of evidence are conducted on the MMCV dataset, including text, images, and tables. The experimental results are shown in Figure 3. From the results, it can be seen that DACLR has the best retrieval performance for text evidence, while its retrieval performance for image evidence is relatively weak. This is consistent with the results of the comparative experiments. Thanks to the powerful semantic understanding ability of the large language model, DACLR's retrieval ability for table evidence has approached that of text evidence. Both have achieved a Recall@20 of 70+\%, a MRR@20 of 69+\%, and an NDCG@20 of 66+\%.

\subsection{Performance on claim verification}

The original intention of the DACLR is to select the most relevant evidence from the multi-modal evidence candidate set to assist the downstream model in claim verification. Existing studies have proved the importance of the quality of the retrieved evidence for claim verification. Based on this, further experiments are conducted to prove that the evidence extracted by DACLR can be better utilized by the downstream model. Using the classic RAG architecture, the Top-10 evidence retrieved by DACLR is used for verification. GPT-4o\cite{openai2024gpt4ocard} and Gemini 1.5 Flash\cite{geminiteam2024gemini15unlockingmultimodal} are  selected as base model, and LLaVA-V1.5-7B\cite{liu2024improvedbaselinesvisualinstruction}, which is an open-source multimodal model representative, is also selected. The experimental results are shown in Table 3. The results indicate that the MLLM using the evidence retrieved by DACLR achieved the best performance indicators. This shows the contribution of DACLR to the entire fact checking process, significantly improving the performance of end-to-end fact checking through better evidence retrieval.

\begin{table}[]
\centering
\resizebox{\linewidth}{!}{
\begin{tabular}{lcccccc}
\hline
Method & Precision & Recall & F1  \\
\hline
GPT-4o & 70.39 & 65.74 & 67.99  \\
GPT-4o + MARVEL & 70.82 & 67.25 & 68.99 \\
GPT-4o + CIEA & 71.36 & 68.93 & 70.12  \\
GPT-4o + DACLR & \textbf{74.88} & \textbf{69.43} & \textbf{72.05}  \\
\hline
GEMINI & 71.53 & 66.12 & 68.72  \\
GEMINI + MARVEL & 72.34 & 67.82 & 70.01 \\
GEMINI + CIEA & 72.17 & 68.35 & 70.21  \\
GEMINI + DACLR & \textbf{74.92} & \textbf{70.01} & \textbf{72.38}  \\
\hline
LLaVa & 64.93 & 63.22 & 64.06  \\
LLaVa + MARVEL & 64.97 & 63.27 & 64.11 \\
LLaVa + CIEA & 65.05 & 64.12 & 64.58  \\
LLaVa + DACLR & \textbf{65.07} & \textbf{64.18} & \textbf{64.62}  \\
\hline
\end{tabular}
}
\caption{Performance on claim verification.}
\label{tab:claim_verification}
\end{table}

\subsection{Dynamic Parameters}

\begin{figure}[t]
  \includegraphics[width=\columnwidth]{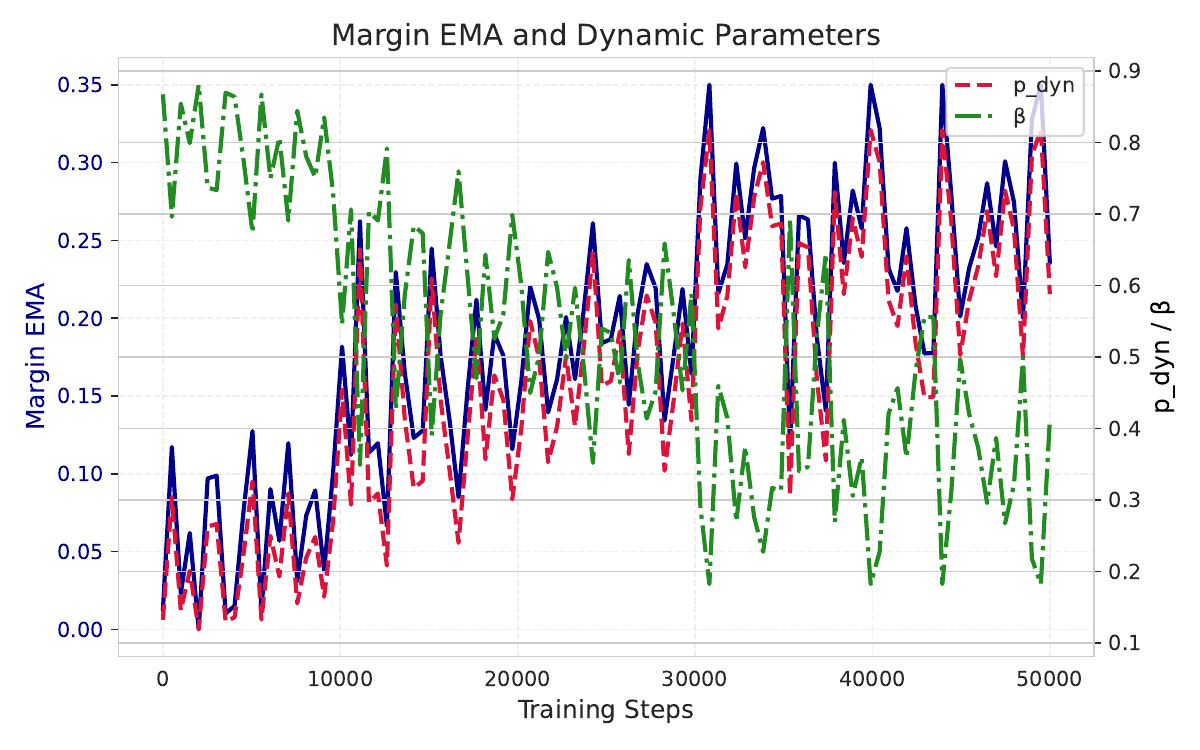}
  \caption{The curve of dynamic parameters during training.}
  \label{fig:experiments}
\end{figure}

During the DACLR training process, the dynamic parameters $p_{dyn}$ and $\beta$ represent the ratio of hard negative samples and event loss, and their changing trends during the training process are shown in Figure 4. As the training progresses, $p_{dyn}$ shows a gradually increasing trend, while $\beta$ is exactly the opposite. Although their values may drop or rise to a certain extent with each negative sample's updating, the overall trend remains unchanged. This also indicates that the model gradually acquires the ability to distinguish positive and negative samples from the perspective of event during the training process. The proportion distribution of negative samples during the training process is shown in Figure 5. As the model's ability increases, the proportion of difficult negative samples in the negative samples also increases, enabling the model to conduct more detailed learning of positive and negative sample distinctions.

\begin{figure}[t]
  \includegraphics[width=\columnwidth]{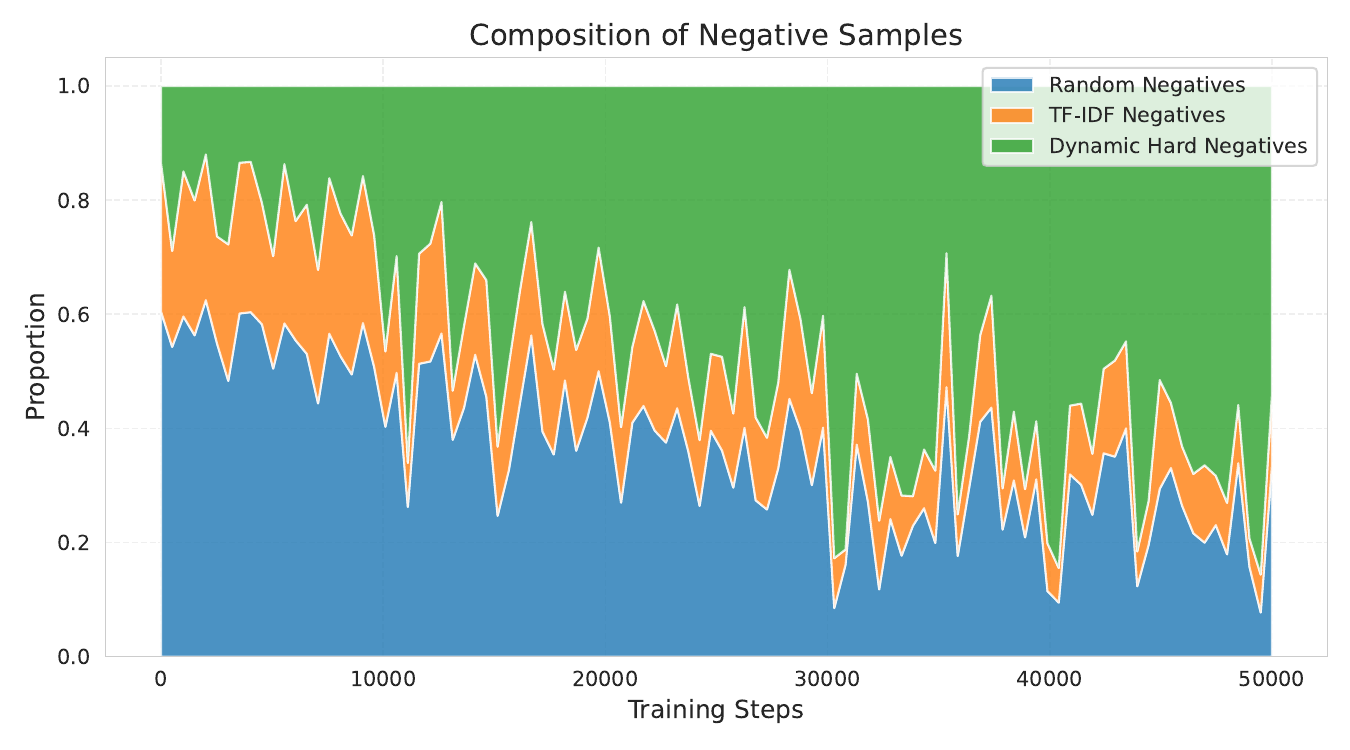}
  \caption{The variation in the proportion of negative samples during training.}
  \label{fig:experiments}
\end{figure}

\subsection{Conclusion}

This paper proposes a multimodal evidence retrieval method named DACLR, which consists of three parts: event summary generation, two-stage evidence ranking, and model training optimization. Firstly, using MLLM, unified event summaries are extracted for claims and multimodal evidence. Then, through the two-stage retrieval, evidence are retrieved according to the claim. In the model training optimization, to enable the model to learn the connection between claims and evidence at the event level, DACLR optimizes the contrastive learning from two aspects: loss function and hard negative sample mining. Two-layer contrastive loss and three-layer hard negative sample sets are designed, and the ratio of the two is dynamically updated based on the intra-batch sample accuracy of the model.
To verify the effectiveness of DACLR in multimodal evidence retrieval, extensive comparative and ablation experiments are conducted. The experimental result demonstrate that DACLR not only outperforms existing general multimodal retrieval methods in multimodal datasets, but also outperforms existing text evidence retrieval methods in pure text evidence retrieval. Moreover, the ablation experiment prove the indispensability of the dynamic loss function and hard negative sample ratio for DACLR. Furthermore, the downstream claim verification experiment also demonstrate that the evidence retrieved by DACLR is more relevant to the claim than that retrieved by existing methods.

\section*{Limitations}

Although the expected results is achieved, DACLR still has several limitations that need to be addressed in future work. The first limitation lies in its dependence on MLLM. As the first and most crucial step of DACLR, the quality of event summaries generated by MLLM has a significant impact on the final retrieval effect. Although the current understanding and generation capabilities of MLLM are quite strong, its unstable output still needs to be resolved. When the output of MLLM contains poor-quality event summaries, it will have a certain impact on DACLR. Another limitation arising from this is the scalability of the modality. DACLR is mainly focused on text and image modality for fact verification. If it is to be expanded to audio or video modalities in the future, it will be limited by the performance of MLLM in that modality.




\bibliography{custom}

\clearpage

\appendix

\section{Algorithm}
\label{sec:appendix}

\begin{algorithm}[H]
\caption{Dynamic Adaptive Contrastive Learning Training}
\begin{algorithmic}[1]
\STATE \textbf{Input}: Claim set $C$, evidence corpus $E$, precomputed fixed negative sample pool, initial model $M_0$
\STATE \textbf{Initialize}: $\bar{\Delta} \leftarrow 0$, $\text{Acc}_{\text{val}} \leftarrow 0$, dynamic negative pool $\mathcal{P}_{\text{dyn}} \leftarrow \emptyset$
\FOR{$t = 1$ to $T$}
    \STATE $B \leftarrow$ sample a batch from training set
    \STATE $\mathbf{h}_c, \mathbf{h}_{e^+}, \mathbf{h}_{\mathcal{N}} \leftarrow M_t(B)$ \hfill \# compute embeddings
    \STATE $\Delta_t \leftarrow$ compute margin following Eq. (1)
    \STATE $\bar{\Delta}_t \leftarrow 0.9\bar{\Delta}_{t-1} + 0.1\Delta_t$
    \IF{$t \bmod U_{\text{eval}} = 0$}
        \STATE $\text{Acc}_{\text{val}} \leftarrow$ retrieval accuracy on validation set
    \ENDIF
    \STATE $\Delta_{\text{mid}} \leftarrow \Delta_{\min} + (\Delta_{\max} - \Delta_{\min}) \cdot \text{Acc}_{\text{val}}$
    \STATE $p_{\text{dyn}} \leftarrow \sigma((\bar{\Delta}_t - \Delta_{\text{mid}})/\tau_s)$, $\beta \leftarrow 1 - p_{\text{dyn}}$
    \STATE Sample negatives: $K \cdot (1 - p_{\text{dyn}})$ from fixed pool, $K \cdot p_{\text{dyn}}$ from $\mathcal{P}_{\text{dyn}}$
    \STATE Compute $\mathcal{L}_{\text{full}}$ and $\mathcal{L}_{\text{unit}}$ (with current $\beta$)
    \STATE $\mathcal{L} \leftarrow \mathcal{L}_{\text{unit}} + \mathcal{L}_{\text{struct}}$
    \STATE Backpropagate and update $M_t \rightarrow M_{t+1}$
    \IF{$t \bmod U_{\text{hard}} = 0$}
        \STATE Mine hard negatives using $M_{t+1}$, update $\mathcal{P}_{\text{dyn}}$
    \ENDIF
\ENDFOR
\STATE \textbf{Output}: Trained retrieval model $M_T$
\end{algorithmic}
\end{algorithm}

Algorithm 1 describes the dynamic adaptive contrastive learning process. For a given claim \(c\) and evidence corpus \(E\), the model maintains both a fixed negative sample pool (precomputed via random and TF-IDF sampling) and a dynamic hard negative pool \(\mathcal{P}_{\text{dyn}}\) updated during training. At each training step \(t\), the model computes embeddings for the claim, positive evidence, and sampled negative evidence. The margin \(\Delta_t\) between positive and negative similarities is calculated following Equation (7), and its exponential moving average \(\bar{\Delta}_t\) is updated to reflect the model's current discriminative capability. The validation retrieval accuracy \(\text{Acc}_{\text{val}}\) is periodically evaluated to adaptively calibrate the threshold \(\Delta_{\text{mid}}\). The dynamic negative sampling ratio \(p_{\text{dyn}}\) and the structure loss weight \(\beta\) are then derived via a sigmoid mapping: \(p_{\text{dyn}} = \sigma((\bar{\Delta}_t - \Delta_{\text{mid}})/\tau_s)\) and \(\beta = 1 - p_{\text{dyn}}\). For each claim, \(K \cdot (1 - p_{\text{dyn}})\) negatives are drawn from the fixed pool while \(K \cdot p_{\text{dyn}}\) are drawn from the dynamic pool \(\mathcal{P}_{\text{dyn}}\). The total loss \(\mathcal{L} = \mathcal{L}_{\text{full}} + \mathcal{L}_{\text{unit}}\), where \(\mathcal{L}_{\text{unit}} = \beta \cdot \mathcal{L}_{\text{sent}} + (1-\beta) \cdot \mathcal{L}_{\text{struct}}\), is used to update the model via backpropagation. Additionally, every \(U_{\text{hard}}\) steps, the current model mines new hard negatives to refresh \(\mathcal{P}_{\text{dyn}}\). As training progresses, higher margin and validation accuracy lead to increased \(p_{\text{dyn}}\) and decreased \(\beta\), enabling the model to gradually shift from easy, fine-grained features to hard, global event‑level matching.

\section{Dataset Statistics}

The data statistics of all the datasets used in the experiment is shown in Table 4.

\begin{table}
\renewcommand{\arraystretch}{1.5}
  \centering
  \resizebox{\linewidth}{!}{
  \begin{tabular}{ccccc}
    \toprule[1.2pt]
       &  FEVER  & FEVEROUS & MMCV & Mocheg  \\
       \midrule
               
    Supported   &  86,701  & 49,115 & 7,361 & 5,144  \\

 Refuted &  36,441  & 33,669 & 8,208 &   5,855   \\

  NEI & 42,305 &  4,242 & - &   4,602 \\
  \midrule
               
    Text   &  -  & - & 11,815 & 33,880  \\

 Images &  -  & - & 6,073 &   12,112   \\

  Tables & - & - & 8,962 &   - \\

  \bottomrule

  \end{tabular}
  }
  \caption{\label{citation-guide}
    This table shows the dataset statistics. 
  }
\end{table}

\section{Implementation Details}
\label{sec:appendix}

In the data preprocessing stage, 150 pages per claim are retrieved by TFIDF and BM25 separately and merged together by ensemble reranking (Dwork et al., 2001) . Furthermore, we selected 30 candidate evidence from each selected documents.

The hyperparameters used in the training are as shown in Table 5.

\begin{table}

\centering

\begin{tabular}{cc}
\toprule
\textbf{Parameter name} &  \textbf{Parameter value} \\
\midrule
epoch &	40 \\
batch size &	32\\
learning rate &	1e-5\\

\bottomrule
\end{tabular}

\caption{The hyperparameters involved in the training process.}\label{tab1}
\end{table}

\section{Preliminary Experiment}
\label{sec:appendix}

We evaluate the existing general multimodal retrieval methods on the MMCV dataset. The results shows that the performance of the mainstream methods has declined to varying degrees on the multimodal fact-checking dataset compared to the general multimodal retrieval dataset, as shown in Table 6. This further supports the viewpoint we present in the second section: The general multimodal retrieval methods cannot be directly applied to evidence-based fact-checking retrieval.

\begin{table}[htbp]

\centering

\label{tab:comparison}
\resizebox{\linewidth}{!}{
\begin{tabular}{@{}lcccccc@{}}
\toprule
\multirow{2}{*}{Method} & \multicolumn{3}{c}{WebQA} & \multicolumn{3}{c}{MMCV} \\
\cmidrule(lr){2-4} \cmidrule(lr){5-7}
                        & MRR@20 & NDCG@20 & Recall@20 & MRR@20 & NDCG@20 & Recall@20 \\
\midrule
BM25                   & 22.80  & 25.41   & 46.27     & 21.45  & 24.24   & 45.44     \\
VinVL-DPR              & 38.74  & 37.79   & 53.89     & 34.74  & 35.66   & 50.12     \\
CLIP-DPR        & 49.34  & 49.11   & 69.84     & 39.50  & 36.23   & 55.37     \\
UniVL-DPR              & 62.96  & 61.22   & 80.37     & 41.24  & 40.27   & 55.24     \\
T5-ANCE                & 64.40  & 64.02   & 83.81     & 46.34  & 47.25   & 61.35     \\
MARVEL          & 65.67  & 65.00   & 84.32     & 50.97  & 48.06   & 63.46     \\
CIEA           & 66.41  & 65.85   & 85.43     & 52.67  & 45.69   & 67.70     \\
\bottomrule
\end{tabular}
}
\caption{Performance comparison on WebQA and MMCV datasets. }
\end{table}

\section{Prompt}
\label{sec:appendix}

The prompt template of MLLM used in the "Event Summary Generation" section is shown in Figure 4.

\begin{figure}[t]
  \includegraphics[width=\columnwidth]{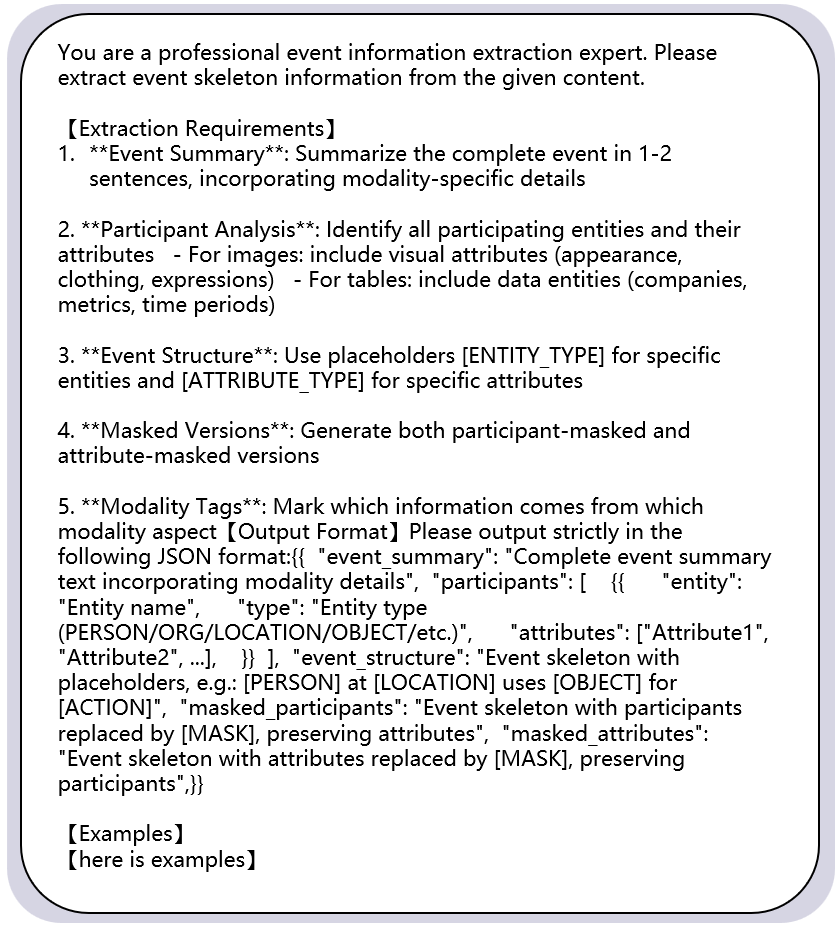}
  \caption{The used prompt template}
  \label{fig:experiments}
\end{figure}

\end{document}